\documentclass[a4paper,11pt]{article}
\pdfoutput=1 

\usepackage{jinstpub} 
\usepackage{subfigure}
\usepackage{booktabs}
\usepackage{hyperref}

\title{\boldmath Potential Advantages of Peak Picking Multi-Voltage Threshold Digitizer in Energy Determination in Radiation Measurement}

\author[a,b]{Kezhang Zhu,}
\author[a]{Junhua Mei,}
\author[a]{Yuming Su,}
\author[a]{Pingping Dai,}
\author[a]{Nicola D'Ascenzo,}
\author[a]{Hao Wang,}
\author[a,c]{Peng Xiao,}
\author[a,d,1]{Lin Wan,\note{Corresponding author.}}
\author[a,c,1]{and Qingguo Xie,}

\affiliation[a]{College of Life Science and Technology, Huazhong University of Science and Technology,\\Wuhan, Hubei, China}
\affiliation[b]{Tongji Hospital, Tongji Medical College, Huazhong University of Science and Technology,\\Wuhan, Hubei, China}
\affiliation[c]{Wuhan National Laboratory for Optoelectronics,\\Wuhan, Hubei, China}
\affiliation[d]{School of Software Engineering, Huazhong University of Science and Technology,\\Wuhan, Hubei, China}

\emailAdd{wanlin@mail.hust.edu.cn (L. Wan),qgxie@mail.hust.edu.cn (Q. Xie)}

\abstract{The Multi-voltage Threshold (MVT) method, which samples the signal by certain reference voltages, has been well developed as being adopted in pre-clinical and clinical digital positron emission tomography(PET) system. To improve its energy measurement performance, we propose a Peak Picking MVT(PP-MVT) Digitizer in this paper. Firstly, a sampled Peak Point(the highest point in pulse signal), which carries the values of amplitude feature voltage and
amplitude arriving time, is added to traditional MVT with a simple peak sampling circuit. Secondly, an amplitude deviation statistical analysis, which compares the energy deviation of various reconstruction models, is used to select adaptive reconstruction models for signal pulses with different amplitudes. After processing 30,000 randomly-chosen pulses sampled by the oscilloscope with a $^{22}Na$ point source, our method achieves an energy resolution of 17.50\% within a 450-650 KeV energy window, which is 2.44\% better than the result of traditional MVT with same thresholds; and we get a count number at 15225 in the same energy window while the result of MVT is at 14678. When the PP-MVT involves less thresholds than traditional MVT, the advantages of better energy resolution and larger count number can still be maintained, which shows the robustness and the flexibility of PP-MVT Digitizer. This improved method indicates that adding feature peak information could improve the performance on signal sampling and reconstruction, which canbe proved by the better performance in energy determination in radiation measurement.}


\arxivnumber{3601876} 

\proceeding{N$^{\text{th}}$ Workshop on X\\
  when\\
  where}

\begin{document}
\maketitle
\flushbottom

\section{Introduction}
\label{sec:intro}

The detection of energetic particles has caught much attention in many areas like industrial radiography~\cite{1}, cosmic radiation detection~\cite{2}, clinical radiotherapy~\cite{3}, particle physics~\cite{4} and positron emission tomography(PET)~\cite{5,6,7} technology. Accurate measurement of the radiation pulse energy is of great importance.

The Multi-voltage Threshold (MVT) method has been well developed as being adopted in preclinical, clinical and application specific digital PET system already~\cite{8,9,10,11}. Traditional MVT method uses several certain voltage thresholds to sample the signal, and usually reconstructs the signal according to some specific pulse models~\cite{12,13} MVT method, which digitalizes the sampling process and avoids fast and high-cost analog-to-digital converters(ADCs), has shown its great advantages as lowering the cost, enhancing the count rate greatly~\cite{14,15,16}. 

Traditional MVT method records the arriving time of the pulse signal when it reaches several preset voltage thresholds to get a set of sampling points, and then uses a specific ideal pulse model to reconstruct the pulse signal via acquired sampling points~\cite{17, 18}. Fixed models and limited thresholds sometimes lead to relatively unsatisfactory energy measurement performance when applied to the radiation detection. Traditional MVT detector has an energy resolution of around 16\% at 511KeV~\cite{9}, while new detectors using sampling methods other than MVT can achieve an energy resolution of 13.01\%~\cite{19}. Methods sharing similar concept with MVT such as time over dynamic threshold (TODT) method and time over threshold (TOT) method~\cite{20, 21}, have achieved the energy resolution of 12.54\%~\cite{22}. Factors like the choice of scintillation crystal and the energy window used for calculating the energy resolution are not same in the work above, thus the results above indeed have limited reference value. However, the traditional MVT could still be thought as some times sacrificing its accuracy on energy determination in some degree. Former works on improving the detector performance focused on physical level such as SiPM (silicon photomultiplier) arrangement~\cite{23} and detector design~\cite{24}. For the traditional MVT detectors, the concept which uses several thresholds and one fixed model is changed in this work.

In this work, we propose a Peak Picking MVT digitizer(also called as Peak Picking MVT or abbreviated as PP-MVT) to elevate MVT's accuracy on the energy determination in radiation measurement while maintaining the high count rate and low cost advantages as possible. In PP-MVT, a peak point (the highest point in pulse signal) is added to traditional sampling points; and multiple adaptive models, rather than single specific model which is fixed for all pulses, are used to reconstruct the pulses of different peak voltages.

To sample the peak point, we give out a simple circuit design. The peak point is unique for one certain corresponding pulse signal and is used for pulses classification: pulses are divided into several types according to their amplitude voltages. And adaptive reconstructing models are used for different pulses. Meanwhile, the peak points are also used in the reconstruction process where adaptive pulse models are used. Our method has achieved better energy resolution and larger number of pulses within energy window of interest for a $^{22}Na$ point source, showing potential advantages in energy determination.

The remainder of this paper is organized as follows. In the methods part, we describe our methods in detail: how the peak points are added to the MVT method, how the peak points could lead to adaptive reconstruction models. Following the methods are the results of the PP-MVT when it is adopted in a $^{22}Na$ point source detection, in comparison with the traditional MVT method. PP-MVT's with less thresholds are also considered. The comparison focuses on the energy resolution and count number of the pulses within the energy window of interest in order to explore new method's potential advantages in energy determination in radiation measurement. Finally, conclusions are drawn and discussion is made.

\section{Methods}
\subsection{Experimental Setup}

\begin{figure}[htbp]
\centering
\includegraphics[width=1.0\textwidth,origin=c]{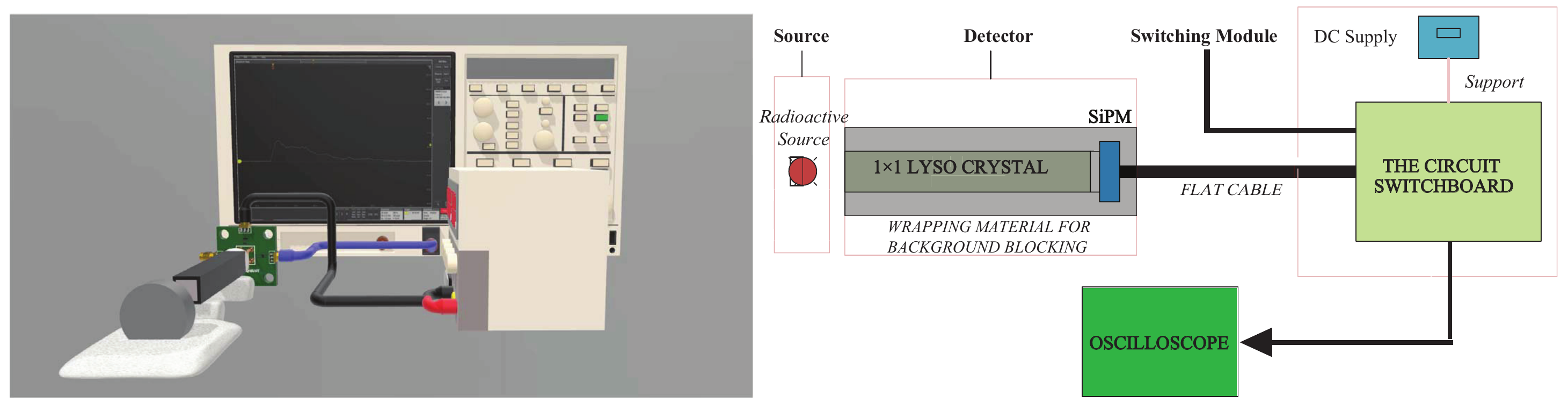}
\caption{\label{fig:i} Experimental Setup for Converting Pulses.}
\label{fig1}
\end{figure}

To convert the original radiation to electrical signal, the high energy gamma photons are first captured and converted in a 3.90*3.90*20 mm$^{3}$ lutetium-yttrium oxy-orthosilicate (LYSO) scintillation crystal~\cite{25,26,27}. Then a silicon photomultiplier (SiPM, Sensl FC 30035) transforms and amplifies the light signal into electric signal~\cite{28,29,30}. SiPM is connected to a Tektronics MSO54 digital oscilloscope (350MHz-2GHz bandwidth, 6.25GS/s sampling rate) through a switch circuit board. The detector is wrapped with blocking material to isolate environment light noise. The experimental setup is shown in Fig.\ref{fig1}.

\subsection{Thresholds}

\begin{figure}[htbp]
\centering 
\subfigure[]{
\label{fig2a}
\includegraphics[width=.46\textwidth,origin=c]{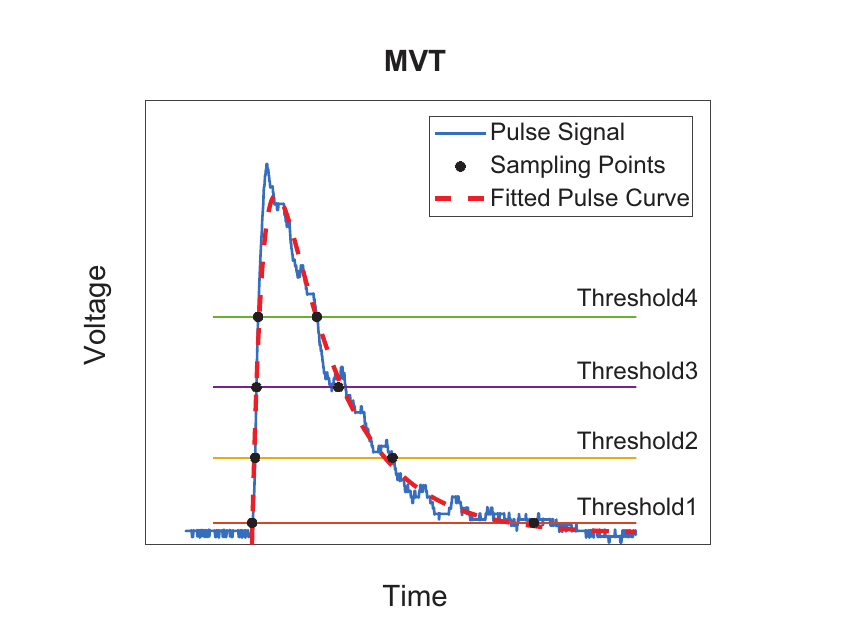}}
\qquad
\subfigure[]{
\label{fig2b}
\includegraphics[width=.46\textwidth,origin=c]{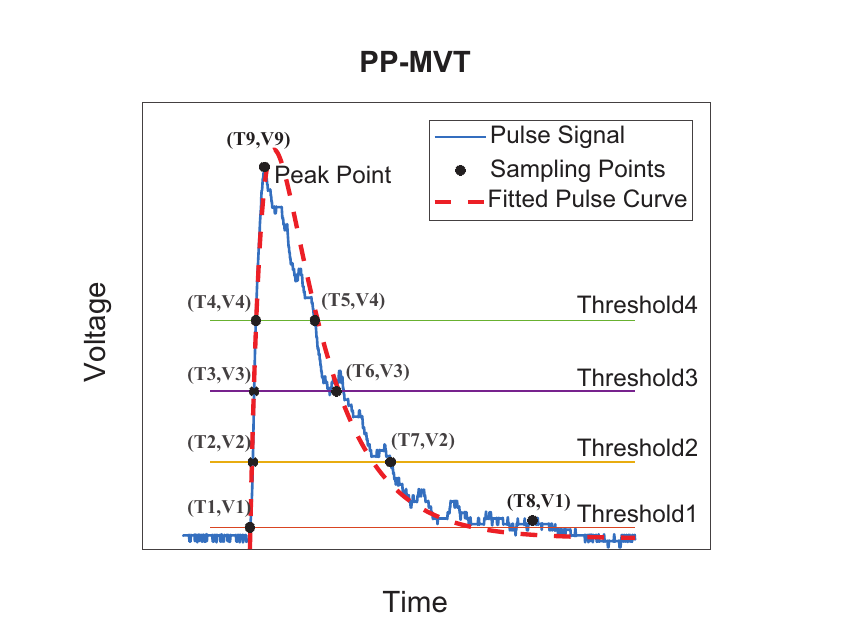}}
\caption{\label{fig:i} (a) The Scheme of traditional MVT. (b) The Scheme of Peak Picking MVT.}
\label{fig2}
\end{figure}

Shown in Fig.\ref{fig2} are the MVT method and the Peak Picking MVT digitizer method. Several voltage thresholds are preset, and they are kept unchanged during the whole sampling process. The selection of the thresholds is important for the performance on energy resolution, and should be explored at first.

\begin{figure}[htbp]
\centering
\includegraphics[width=.6\textwidth,origin=c]{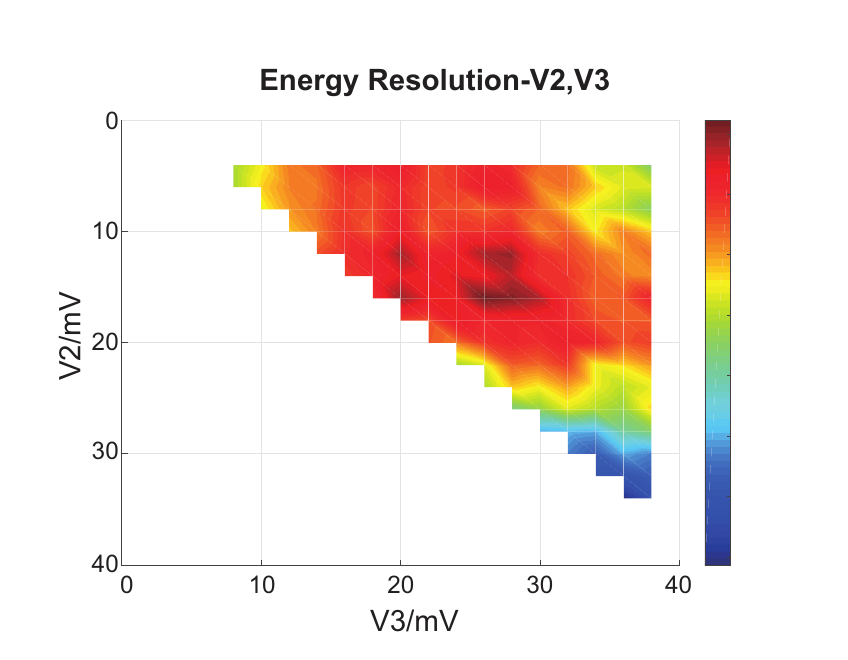}
\caption{\label{fig:i} V2, V3 determination based on energy resolution.}
\label{fig3}
\end{figure}

The number of thresholds is selected as four with the combined consideration of the complexity and accuracy. Four thresholds are labeled as V1, V2, V3 and V4. And in PET, leading edge discriminator (LED), which compares a preset threshold with the pulse signal and records the triggering time when pulse reaches the threshold, is the most widely used time pickoff method for coincidence timing measurement~\cite{32}. Here V1 threshold(the lowest threshold in MVT, or the preset LED threshold) is chosen by time resolution performance, and when setting V1 as 2mV, the best(lowest) time resolution is achieved. The discussion about the time resolution currently is not the focus of this work, thus is left out. V4 is chosen as 40mV to restrict potential invalid noise pulses that usually have low amplitudes.

After V1 and V4 are fixed, V2 and V3 are selected by energy resolution performance shown in Fig.\ref{fig3}, where warmer color represents better(lower) energy resolution. The energy resolution is calculated as the FWHM of the Gaussian fitting curve around the 511KeV photoelectric peak in the energy spectrum. With the result in Fig.\ref{fig3}, [12mV, 26mV] would be an appropriate choice for V2 and V3.

As stated above, the thresholds are chosen as [2mV, 12mV, 26mV, 40mV].

\subsection{Peak Picking Multiple-Voltage-Threshold digitizer}

The improved method adds a peak point to the traditional MVT, and adopts adaptive reconstruction models for the pulses with different peak voltages.

To realize it, two steps are needed:

i) Sample the peak point;

ii) Select pulses' reconstruction models according to their peak voltages.

The whole process is a simulation experiment based on the original pulse data acquired from the OSC to get energy spectrum for result evaluation.

\subsubsection{Overall Framework}

Shown in Fig.\ref{fig2} are the two schemes: the traditional MVT method and the improved Peak Picking MVT digitizer method. 

In the traditional MVT in Fig.\ref{fig2a}, as the pulse is coming, the arriving time when the pulse reaches the threshold would be recorded on both the raising and the falling edges of the pulse. If there are more than one point sampled by a threshold, the assumed arriving time could be determined as their average time. So sampling points of twice the number of the thresholds will be acquired at last. Then the pulse signal will be reconstructed by fitting the sampling points using a certain pulse model. In the Peak Picking MVT in Fig.\ref{fig2b}, a peak point (the highest point in pulse signal) is added to the traditional MVT. The peak point is sampled directly in the simulation experiment, and the design of a simple sampling circuit is given in the next part. To reconstruct the pulses based on several sampled points, adaptive models are used for the pulses owning different peak voltages.

\subsubsection{Peak Point Determination}
\begin{figure}[htbp]
\centering 
\subfigure[]{
\includegraphics[width=0.53\textwidth,origin=c]{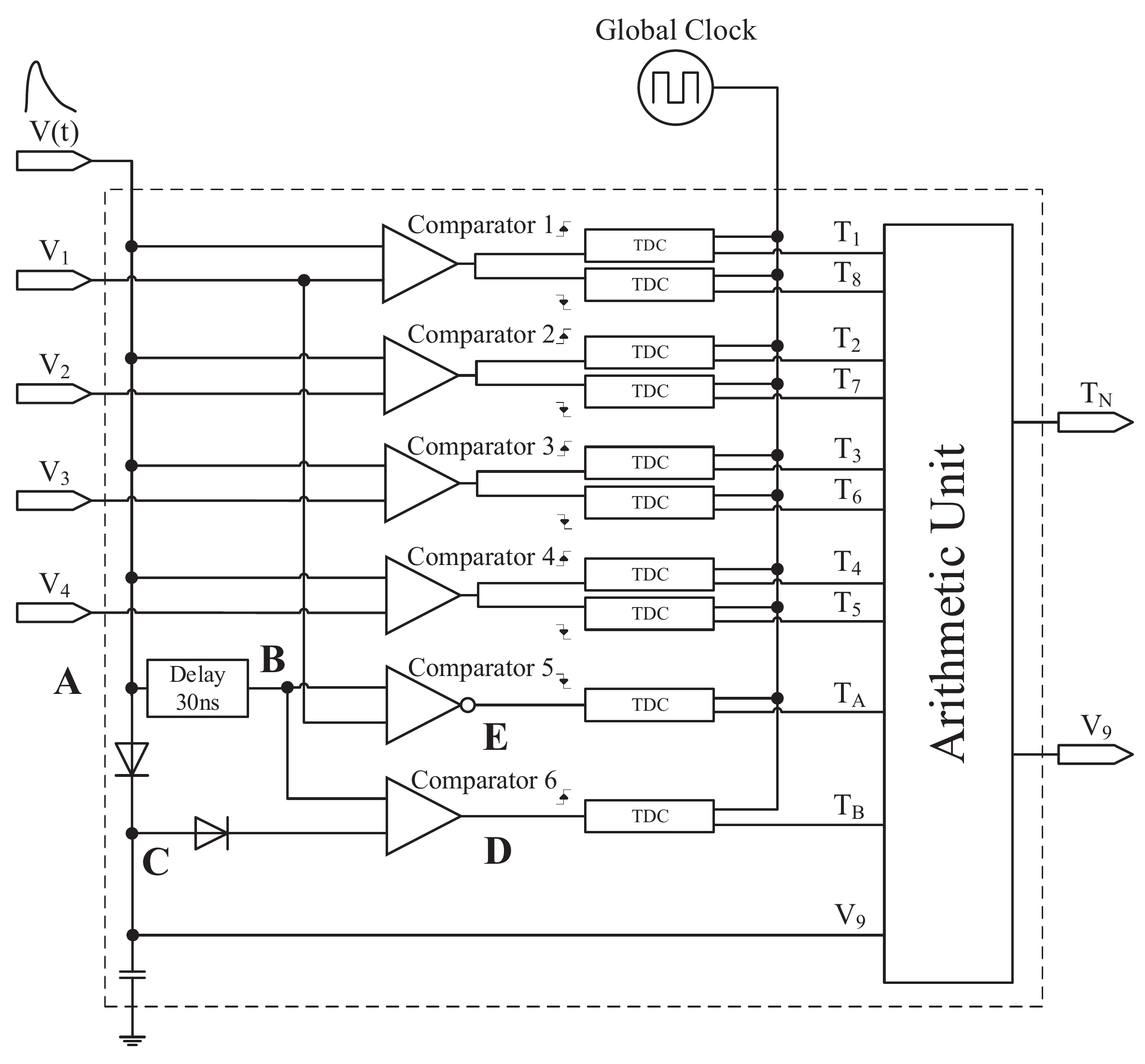}}
\\
\subfigure[]{
\includegraphics[width=0.7\textwidth,origin=c]{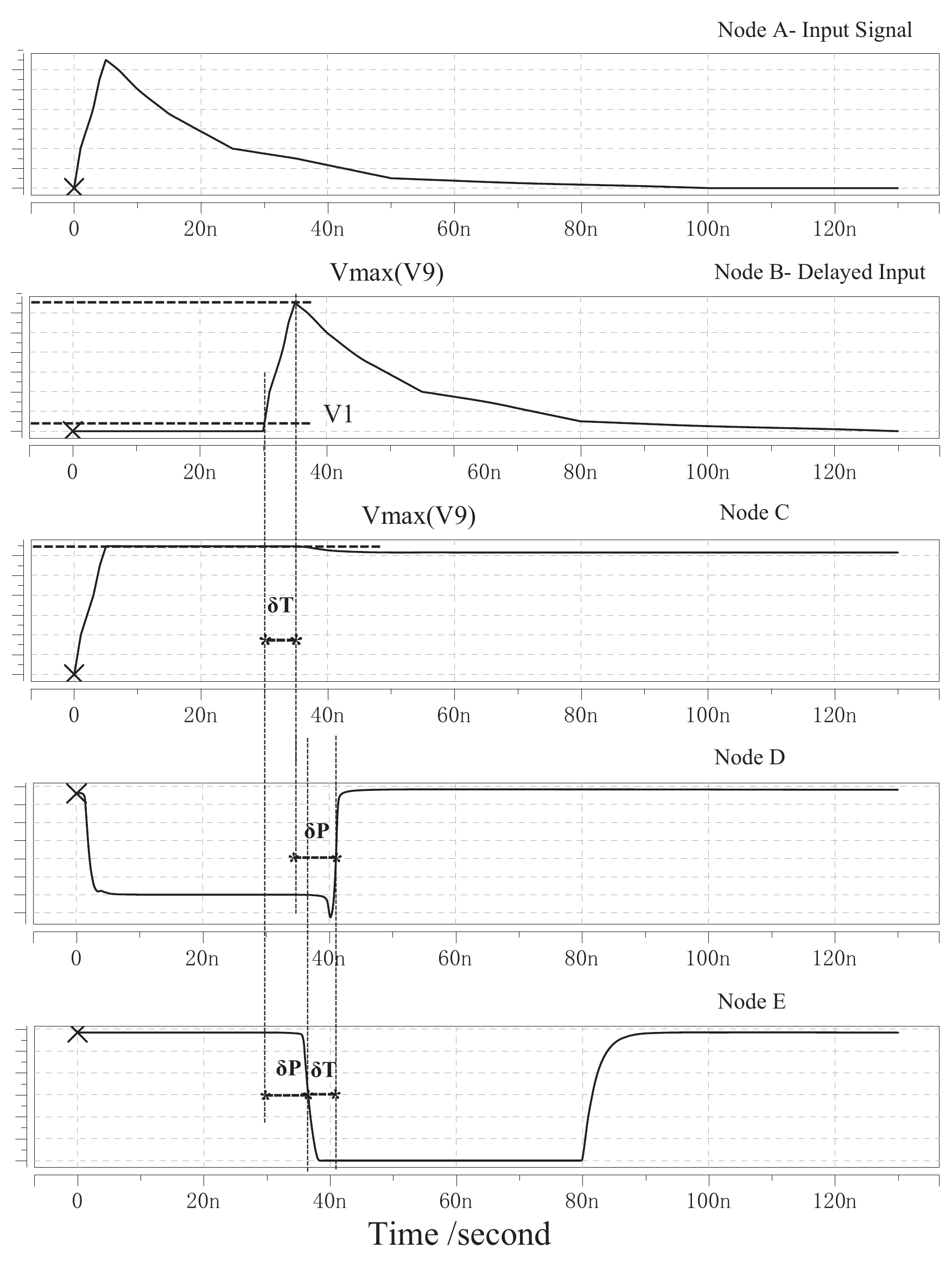}}
\caption{\label{fig:i} (a) The Scheme of traditional MVT. (b) The Scheme of Peak Picking MVT.}
\label{fig4}
\end{figure}

To determine one peak point, we need a time coordinate and a voltage coordinate. As shown in Fig.\ref{fig2b}, the arriving times of the eight points are labeled as T1 to T8, and their voltage values are labeled as V1 to V4. Marking the peak point as (T9, V9). The sampling circuit design and its simulation result by Hspice are presented in Fig.\ref{fig4}. 

The pulse signal V(t) moves through the circuit to leave a peak point voltage V9 and a time interval between the peak point and the first sampling point in MVT. 

The circuit implementation of the traditional MVT has been clearly illustrated in the work of Xi et al[9]. And to additionally sample the peak voltage, the pulse input is fed to node A. The diode conducts before the pulse begins to decline, so the electric potentials between node A and node C should be equal. Once the pulse declines, the diode would be cut off and the peak voltage V9 of the pulse would be kept by a capacitance at node C. To locate the peak at the time axis, a delayed pulse input is introduced to the node B with a proper delay time to make sure that V9 has been gotten at node C. Then the voltage comparator 5 and 6 would individually give out a jump/drop in output end when the delay voltage meets the reference threshold V1 and peak voltage V9. The time interval between the changes in comparators output should be the time interval between the first sampling point in MVT and the peak point. The time counter unit Time to Digital Convertor(TDC) could be used to record the absolute time when the input signal of TDC changes from high to low or low to high.

To simulate the peak sampling circuit by Hspice, a diode with an ultra-small emission coefficient and an ultra-low forward voltage drop is defined and used. The voltage comparators should be of high speed and have a propagation time $\delta$ P of several nanoseconds. The input of the comparators could be amplified to some degree in order to fit the requirements of the comparators. From the simulation result, the circuit works and gives out a peak voltage V9 at node C together with a time interval $\delta$
 T between T9 and T1.

\subsubsection{Model Selection}

\begin{figure}[htbp]
\centering
\subfigure[]{
\label{fig5a}
\includegraphics[width=0.4\textwidth,origin=c]{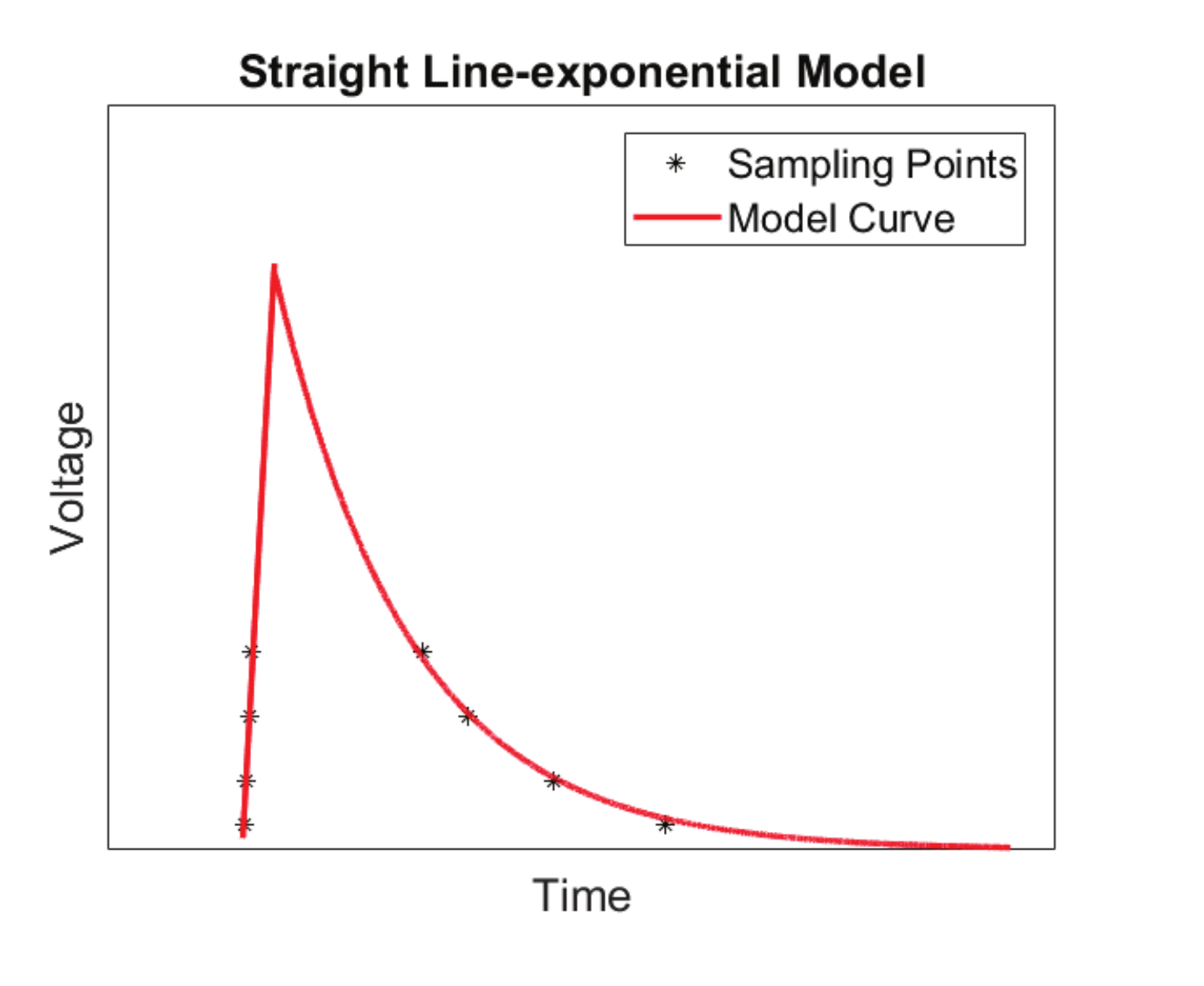}}
\qquad
\subfigure[]{
\label{fig5b}
\includegraphics[width=0.4\textwidth,origin=c]{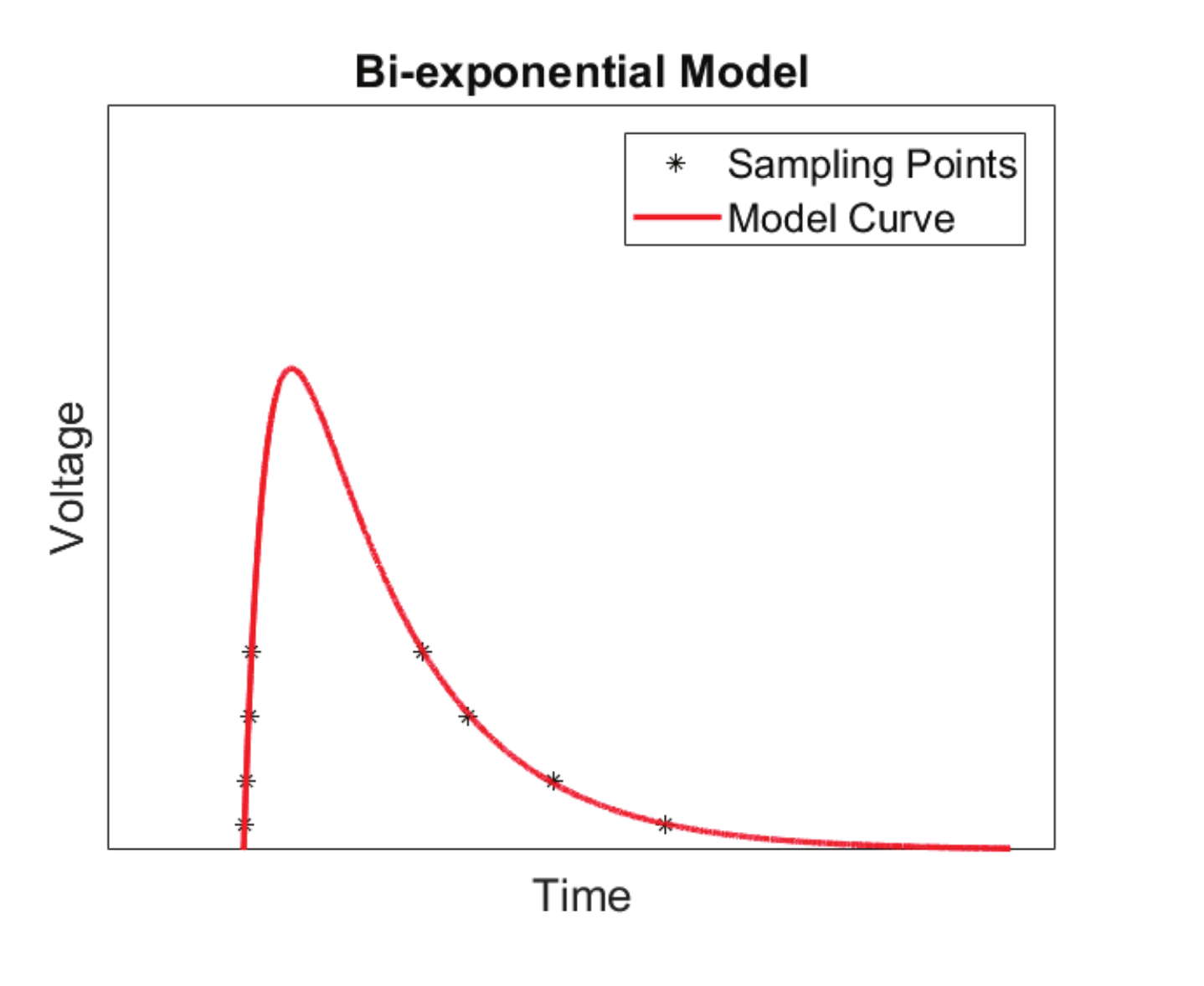}}
\\
\subfigure[]{
\label{fig5c}
\includegraphics[width=0.4\textwidth,origin=c]{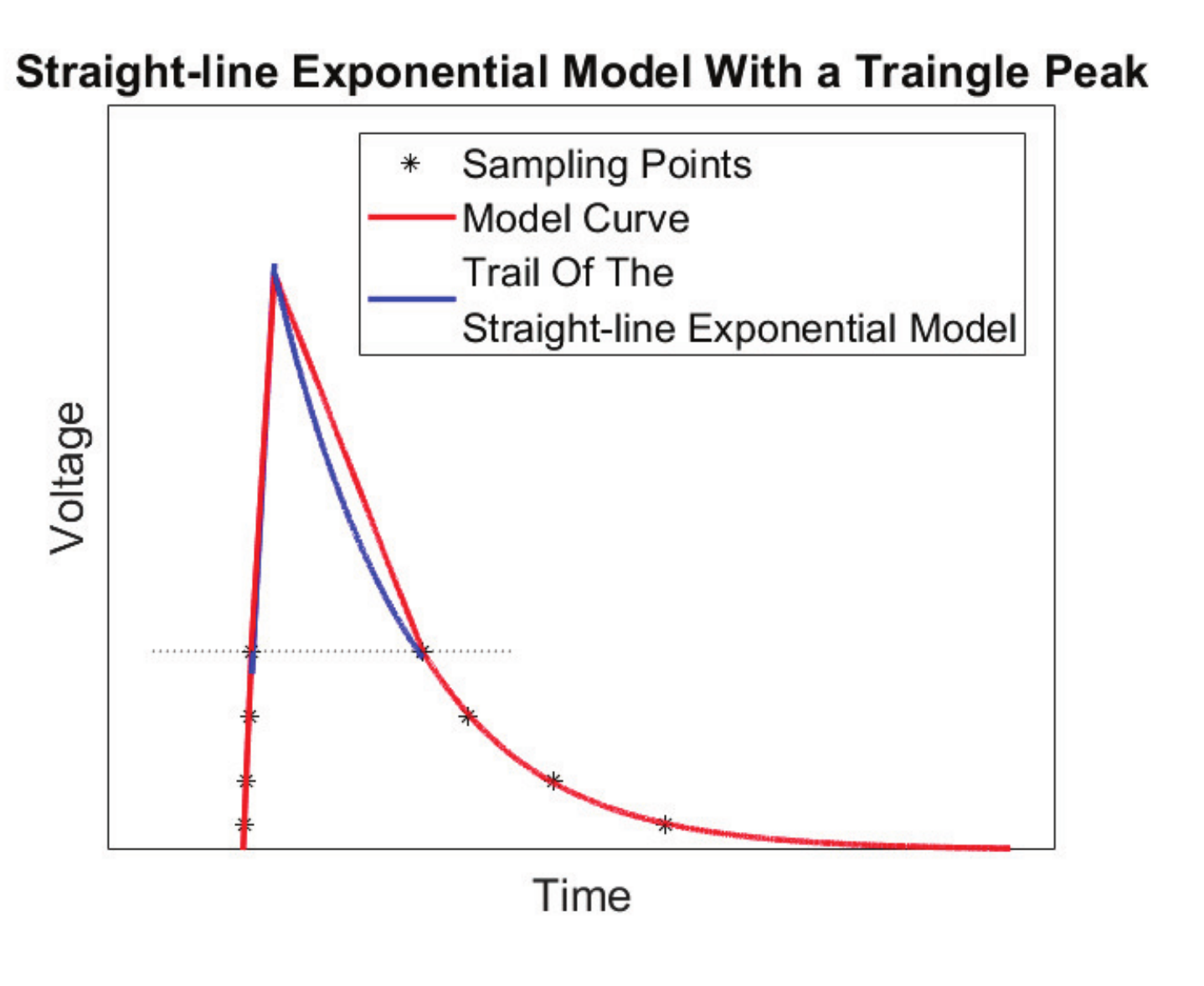}}
\qquad
\subfigure[]{
\label{fig5d}
\includegraphics[width=0.4\textwidth,origin=c]{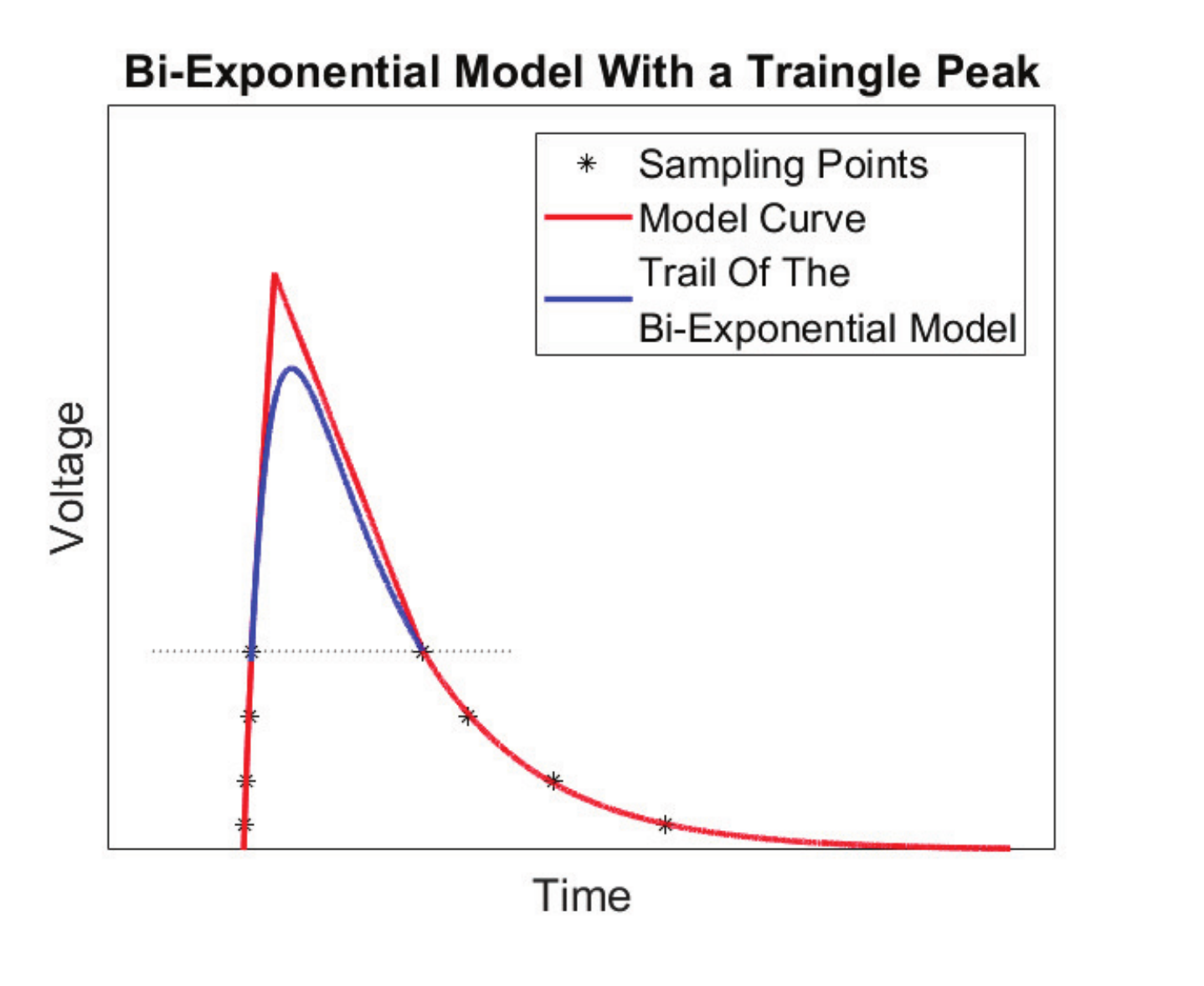}}
\caption{\label{fig:i} Various models for pulse reconstruction. (a) Straight line-exponential model without peak. (b) Bi-exponential model without peak. (c) Straight line-exponential model with a triangle peak. (d) Bi-exponential model with a triangle peak.}
\label{fig5}
\end{figure}

In the traditional MVT, the reconstruction model is fixed for all the pulses, which renders the insufficiency of the flexibility for signals of great random. The introduction of the peak point in PP-MVT makes it possible to solve this problem: peak point functions as a feature for the classification of the pulses, and various models are prepared for different peak voltage intervals. Theoretically speaking, the models available are infinite. And in this work we only consider about several selected ones. Shown in Fig.\ref{fig5} are the four main (not all) models used in our method: Fig.\ref{fig5a} is the straight line exponential model; Fig.\ref{fig5b} is the bi-exponential model; Fig.\ref{fig5c} is the straight line-exponential model with a triangle peak; Fig.\ref{fig5d} is bi-exponential model with a triangle peak. In straight line-exponential model, the raising edge is considered as a straight line and descending edge is considered as an exponential line; and in the bi-exponential model, the pulse is supposed to follow a bi-exponential model trend. And for the peak point, it can either enter the whole general model or form a triangle with the two sampling points below it, while the traditional MVT method in this work adopts bi-exponential model.

\begin{figure}[htbp]
\centering
\subfigure[]{
{\begin{minipage}[t]{0.40\textwidth}
\centering
\includegraphics[width=5.9cm]{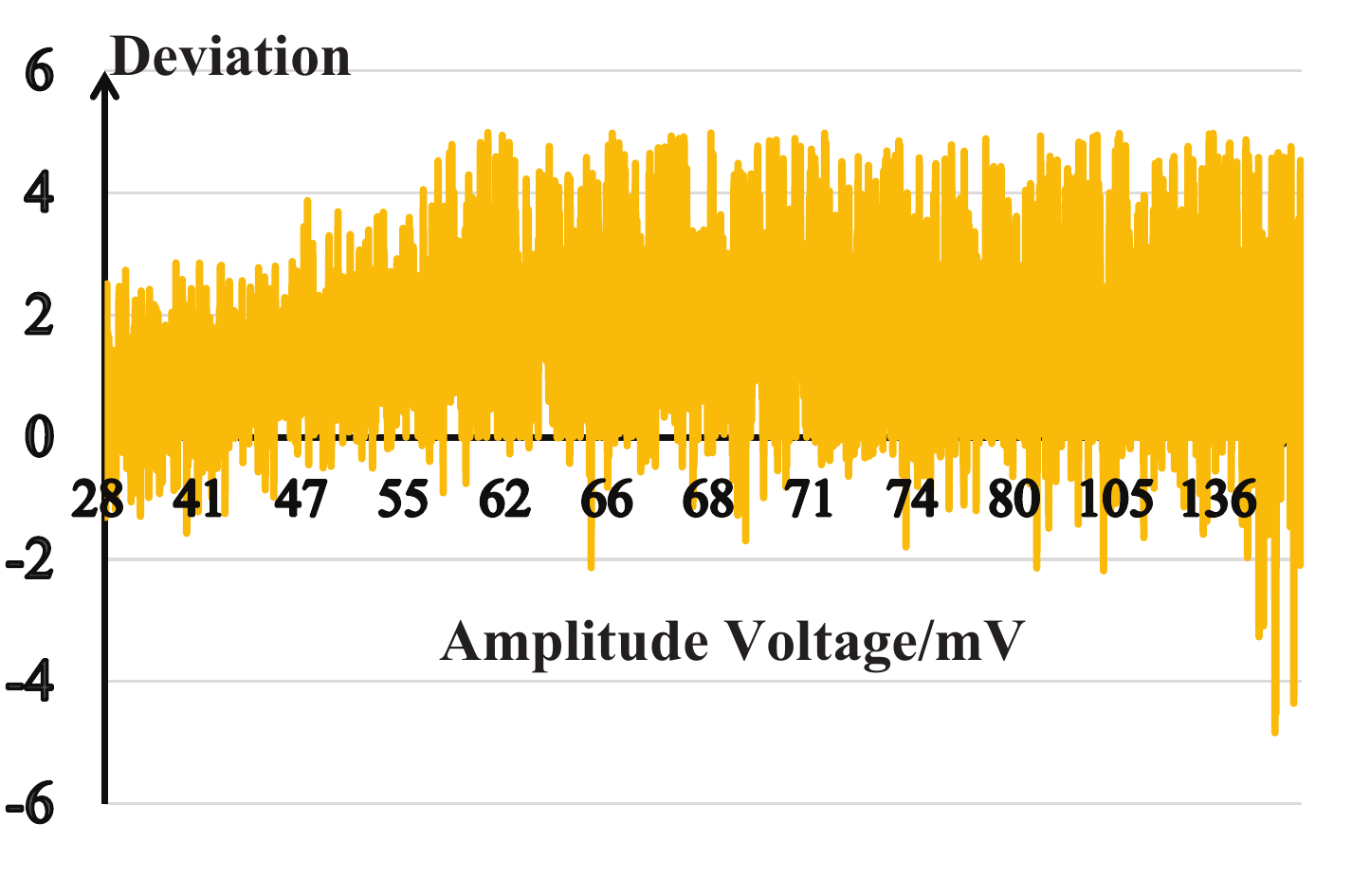}
\end{minipage}}}
\subfigure[b]{
{\begin{minipage}[t]{0.40\textwidth}
\centering
\includegraphics[width=5.9cm]{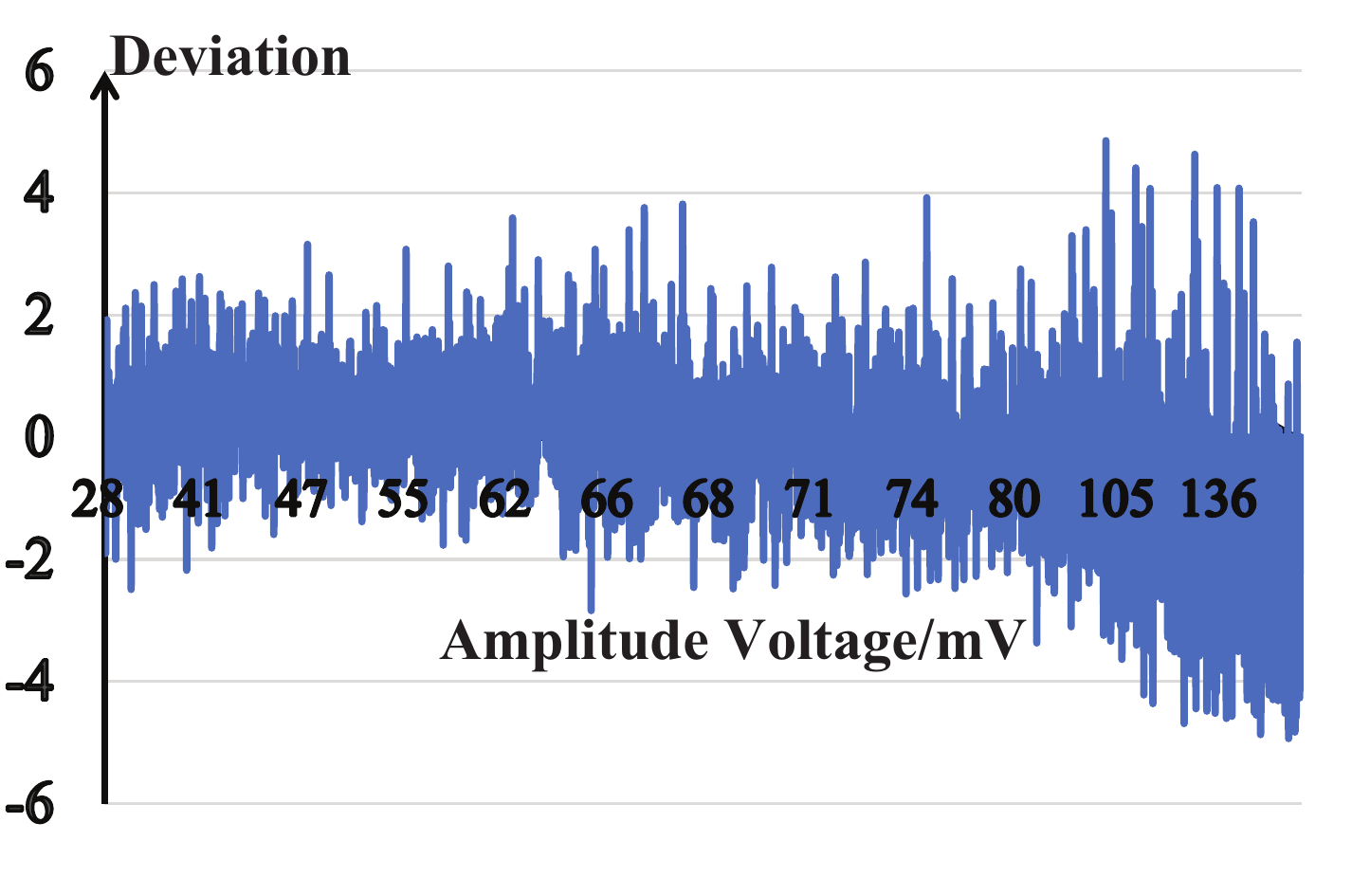}
\end{minipage}}}
\\
\subfigure[]{
{\begin{minipage}[t]{0.40\textwidth}
\centering
\includegraphics[width=5.9cm]{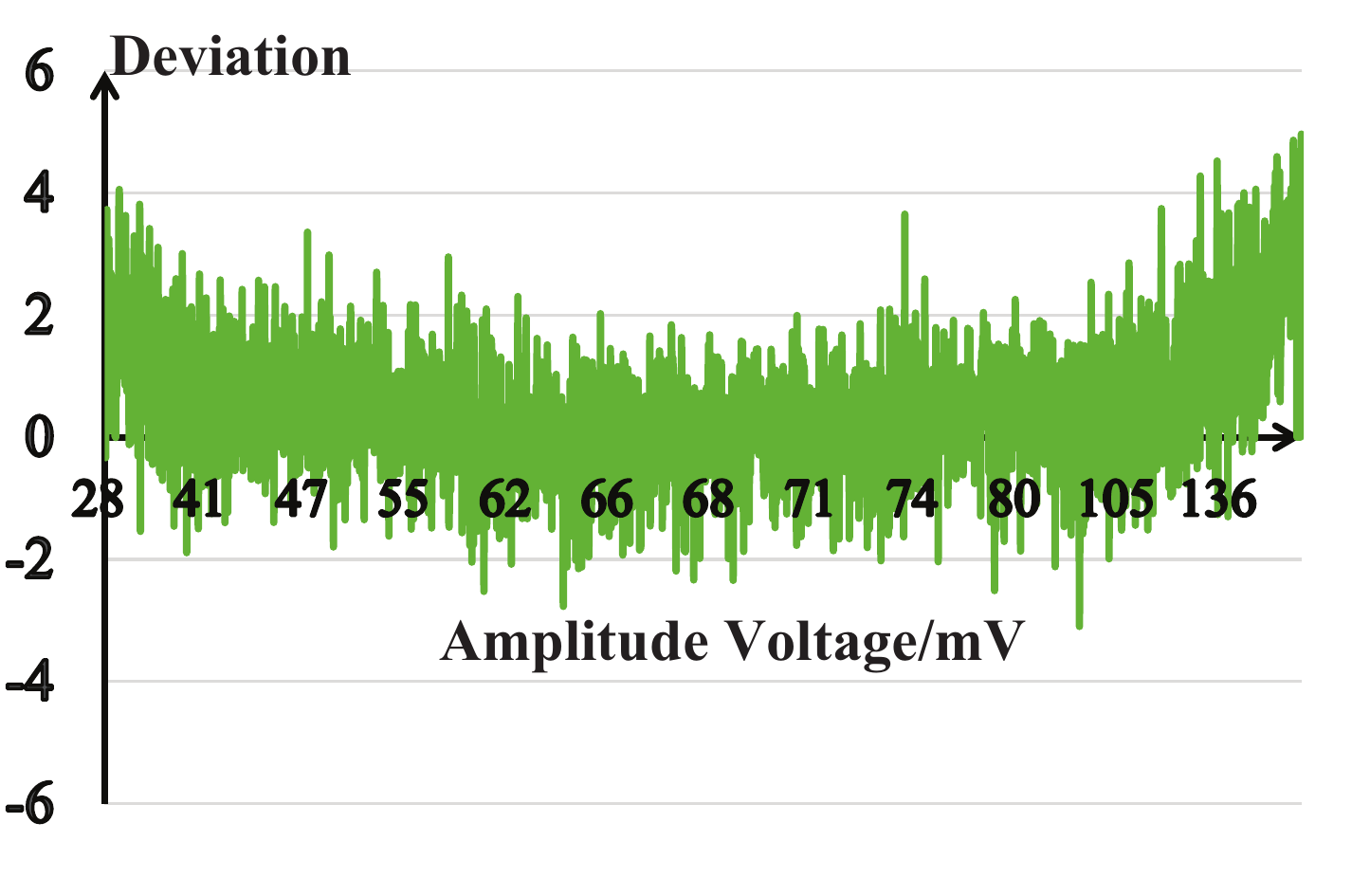}
\end{minipage}}}
\subfigure[]{
{\begin{minipage}[t]{0.40\textwidth}
\centering
\includegraphics[width=5.9cm]{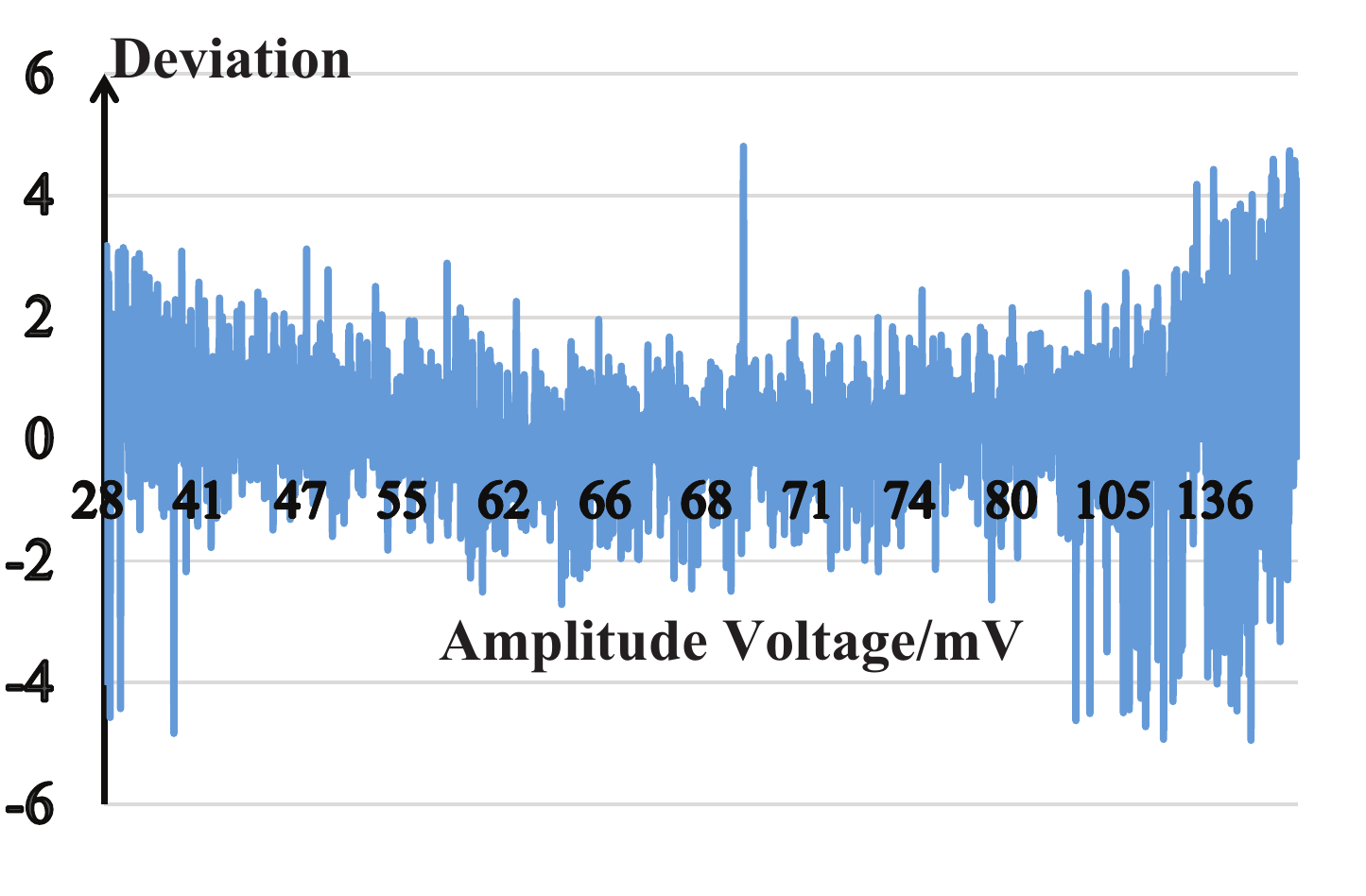}
\end{minipage}}}
\caption{Model deviation analysis for pulses of different peak voltages. (a)Straight line-exponential model without peak. (b) Bi-exponential model without peak. (c) Straight line-exponential model with a triangle peak. (d) Bi-exponential model with a triangle peak.}
\label{fig6}
\end{figure}

Shown in Fig.\ref{fig6}, to determine the proper models adopted for pulses of different voltage intervals, an amplitude deviation statistical analysis is used for the model selection: to select the models, the X coordinate represents the voltage of the peak points of pulses and the Y coordinate represents the deviation between the real energy and the energy calculated by the chosen model. If the deviation values in a peak voltage interval are overall positive for a model, the fitted energy values of the pulses within this voltage interval are overall bigger than the real energy. And it is just the opposite for the negative deviation values. So the proper model for fitting pulses within a peak voltage interval is the model having a low average absolute value of the deviation within this peak voltage region.

To obtain a proper model selection, considering both the accuracy and the complexity of the classification, we make the plan for the four thresholds situation as follows:

i) The peak voltage between 40-50mv: using a bi-exponential model with a triangle peak, while nine points(including the peak point) are involved in the calculation of the bi-exponential curve;

ii) The peak voltage between 50-130mv: using a straight line-exponential model with a triangle peak, while eight points(except the peak point) are involved in the calculation of the straight line-exponential curve;

iii) The peak voltage between 130-155mv: using a biexponential model with a triangle peak, while nine points(including the peak point) are involved in the calculation of the bi-exponential curve;

iv) The peak voltage above 155mv: using a straight line-exponential model, while nine points(including the peak point) are involved in the calculation of the straight line-exponential curve.

\section{Results}

For each pulse signal, the area of the pulse is calculated to obtain its energy. To evaluate the performance in energy determination in radiation measurement of the method proposed in the Section II, the improved Peak Picking MVT digitizer method is compared with the traditional MVT method. And to further appraise the PP-MVT's flexibility and robustness, PP-MVTs with thresholds less than four are also added into the comparison. By considering the results of PP-MVT with less thresholds, the improvement brought by the introduction of the peak point could be further validated.

Following the methods described in section II, the four threshold MVT and PP-MVT should have the thresholds selection [2mV, 12mV, 26mV, 40mV], and the models selection is the plan mentioned above. And getting through the same process stated in the methods part, the best thresholds selection and models selection plans for PP-MVT with three and two thresholds are also determined. Exact plans are easy to get thus not written here. For the one-threshold PP-MVT, only V1(2mV) is maintained and a straight line-exponential model is used for signal reconstruction.

We use one set of 30,000 randomly-chosen pulses sampled by the oscilloscope with a $^{22}Na$ point source for the result evaluation. The comparison references are the energy resolution and the count number of the pulses within the 450-650KeV energy window of interest. 30,000 pulses are sampled within a short time period and the layout of the sampling system should not be changed during the whole data acquisition process to keep the environment a stable as possible.

And the energy values of the pulses should be calibrated before the result evaluation.

\subsection{Energy Calibration: SiPM nonlinearity correction}

The response of the SiPM is not linear with the number of impinging photons. And pulse energy values gotten in the system are not equal to the real energy of the original Gamma quanta due to non-linear SiPM response. The dependence of the output signal of the SiPM from the deposited energy's amount can be described with an exponential function~\cite{31}:
\begin{equation}
\label{eq:x}
V(E)=A\left(1-e^{-\frac{E}{N_{c e l l s} B}}\right)
\end{equation}

\begin{figure}[htbp]
\centering
\includegraphics[width=0.7\textwidth,origin=c]{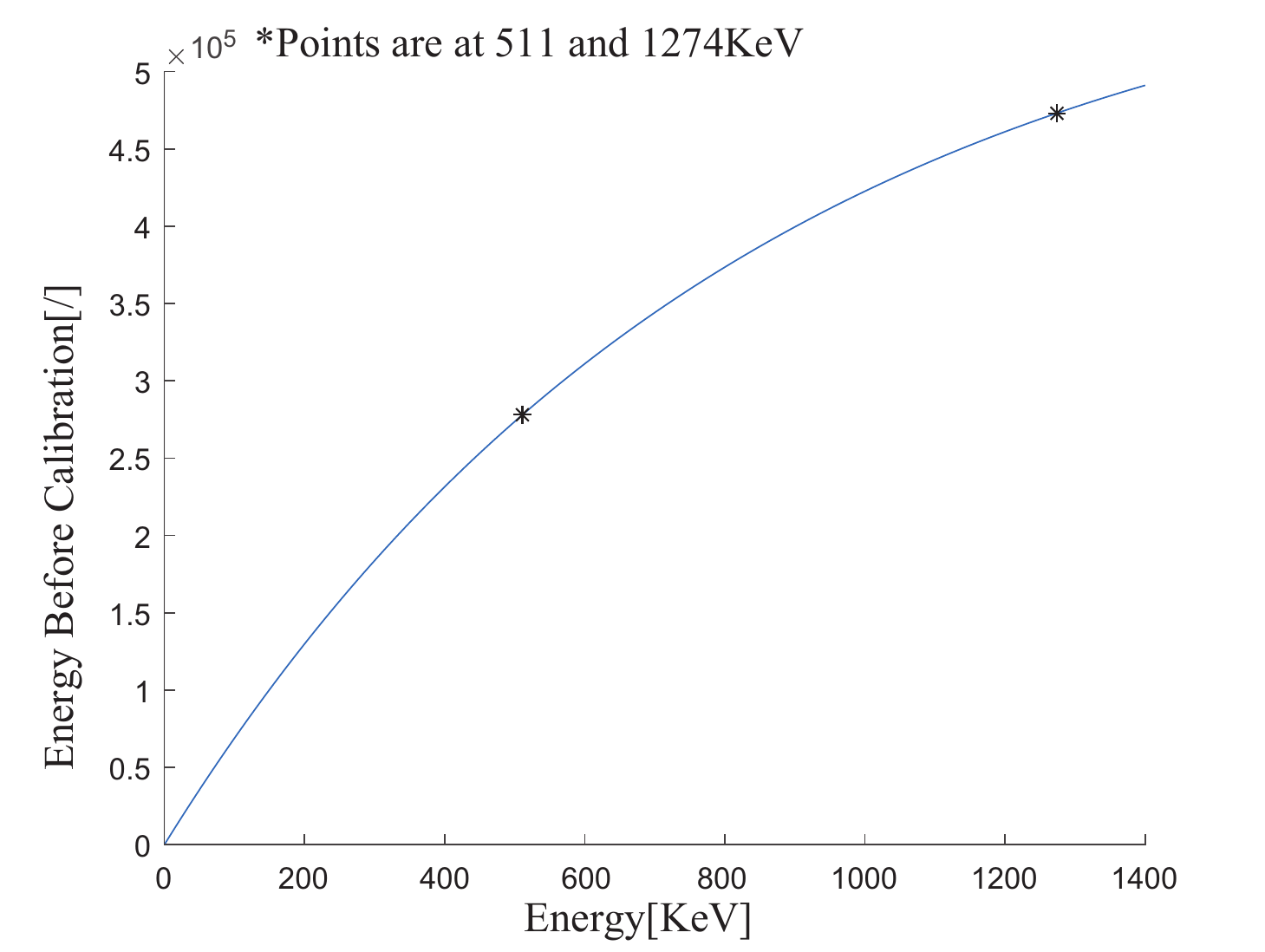}
\caption{\label{fig:i} Experimental Setup for Converting Pulses.}
\label{fig7}
\end{figure}

Where E is the energy value of the Gamma quanta, Ncells is the number of cells in SiPM, and A and B are free parameters. Once the V(E) is known, the real energy of the Gamma quanta could be deprived. One correction curve is shown in Fig.\ref{fig7}, every energy value of the pulse should be calibrated with a certain calibration relationship to recover the real energy.

One correction curve is shown in Fig.\ref{fig7}, every energy value of the pulse should be calibrated with a certain calibration relationship to recover the real energy.

\subsection{Results for single $^{22}Na$ source}

\begin{figure*}[htbp]
\centering
\subfigure[]{
{\begin{minipage}[t]{0.45\textwidth}
\centering
\includegraphics[width=7cm]{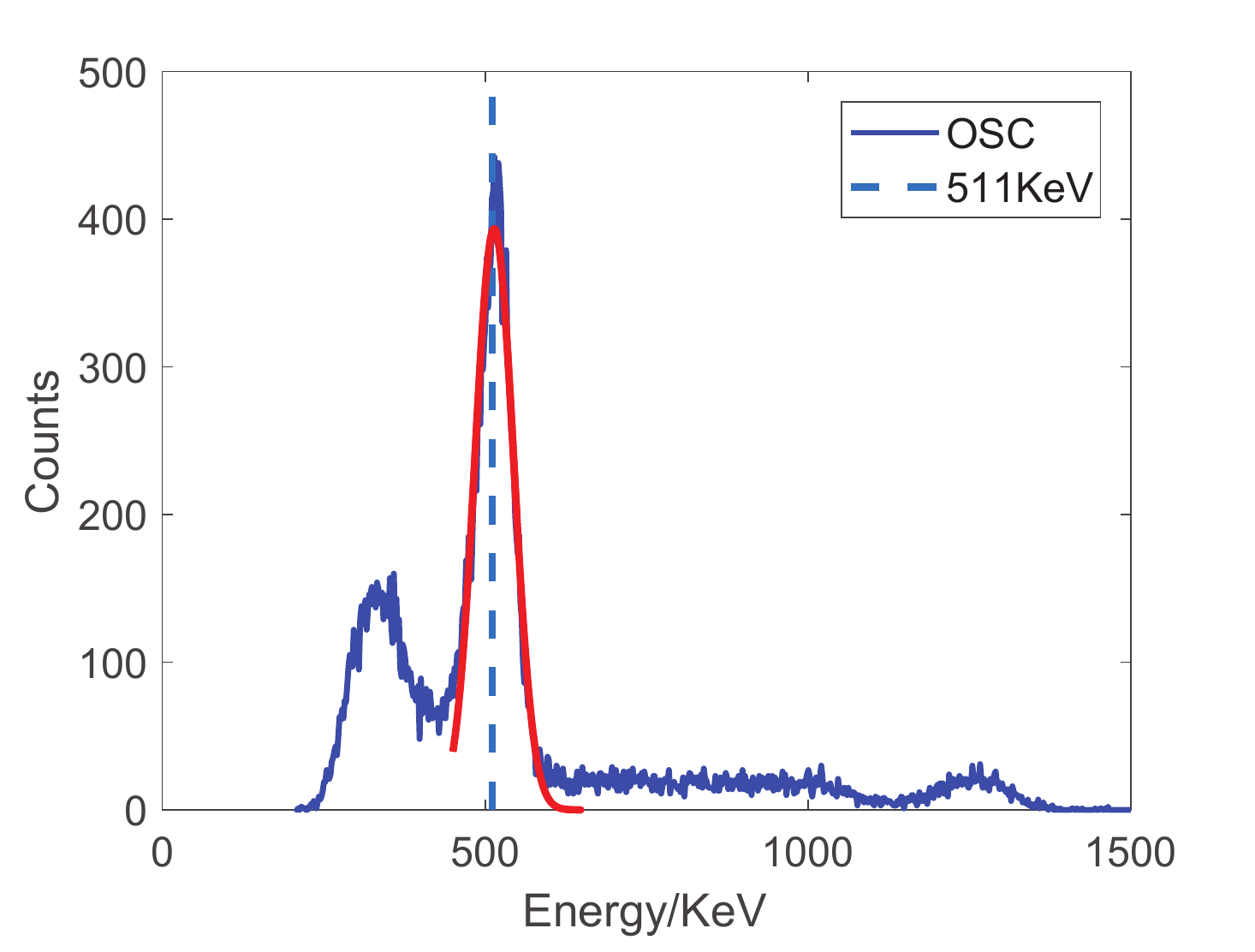}
\end{minipage}}}
\subfigure[]{
{\begin{minipage}[t]{0.45\textwidth}
\centering
\includegraphics[width=7cm]{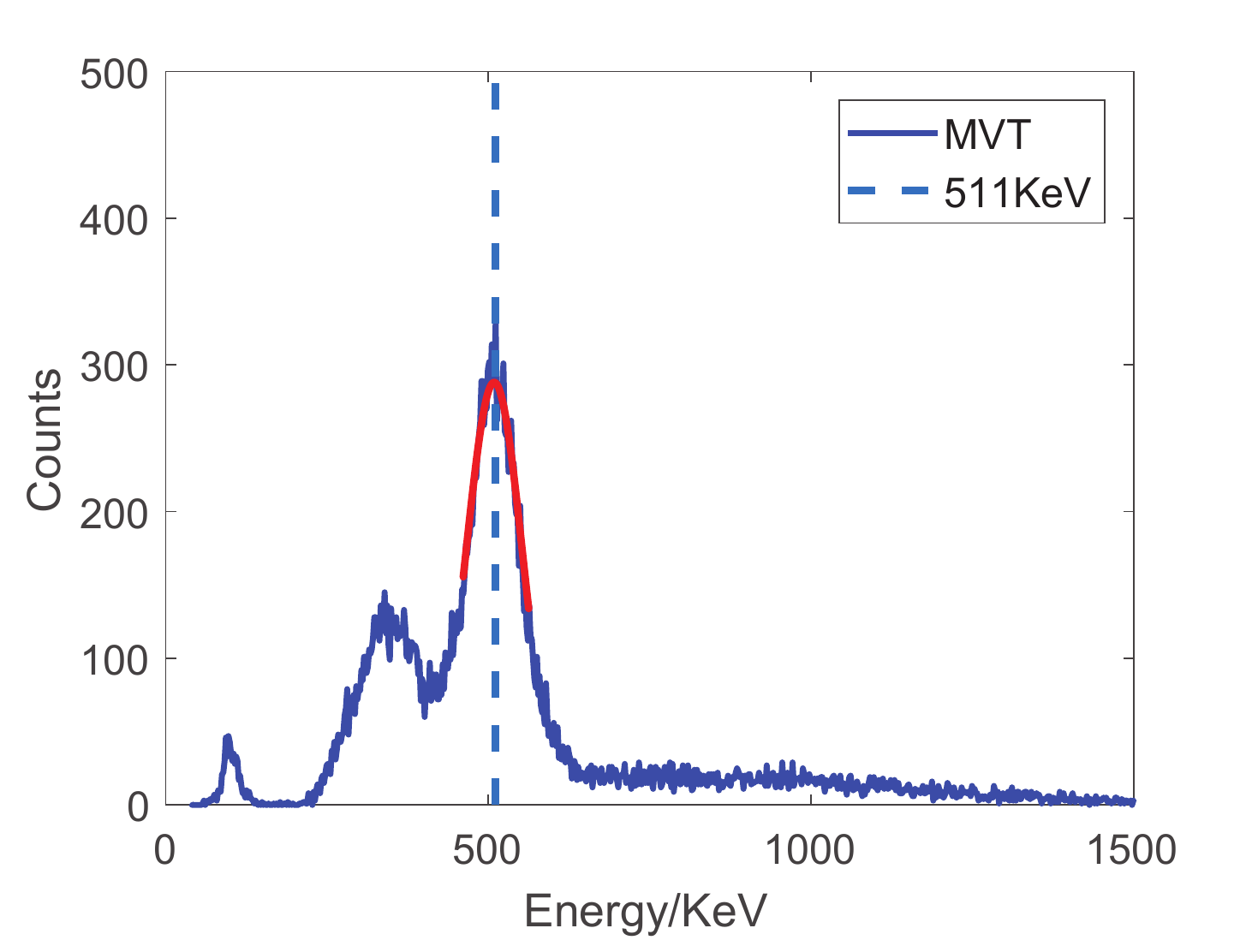}
\end{minipage}}}
\\
\subfigure[]{
{\begin{minipage}[t]{0.45\textwidth}
\centering
\includegraphics[width=7cm]{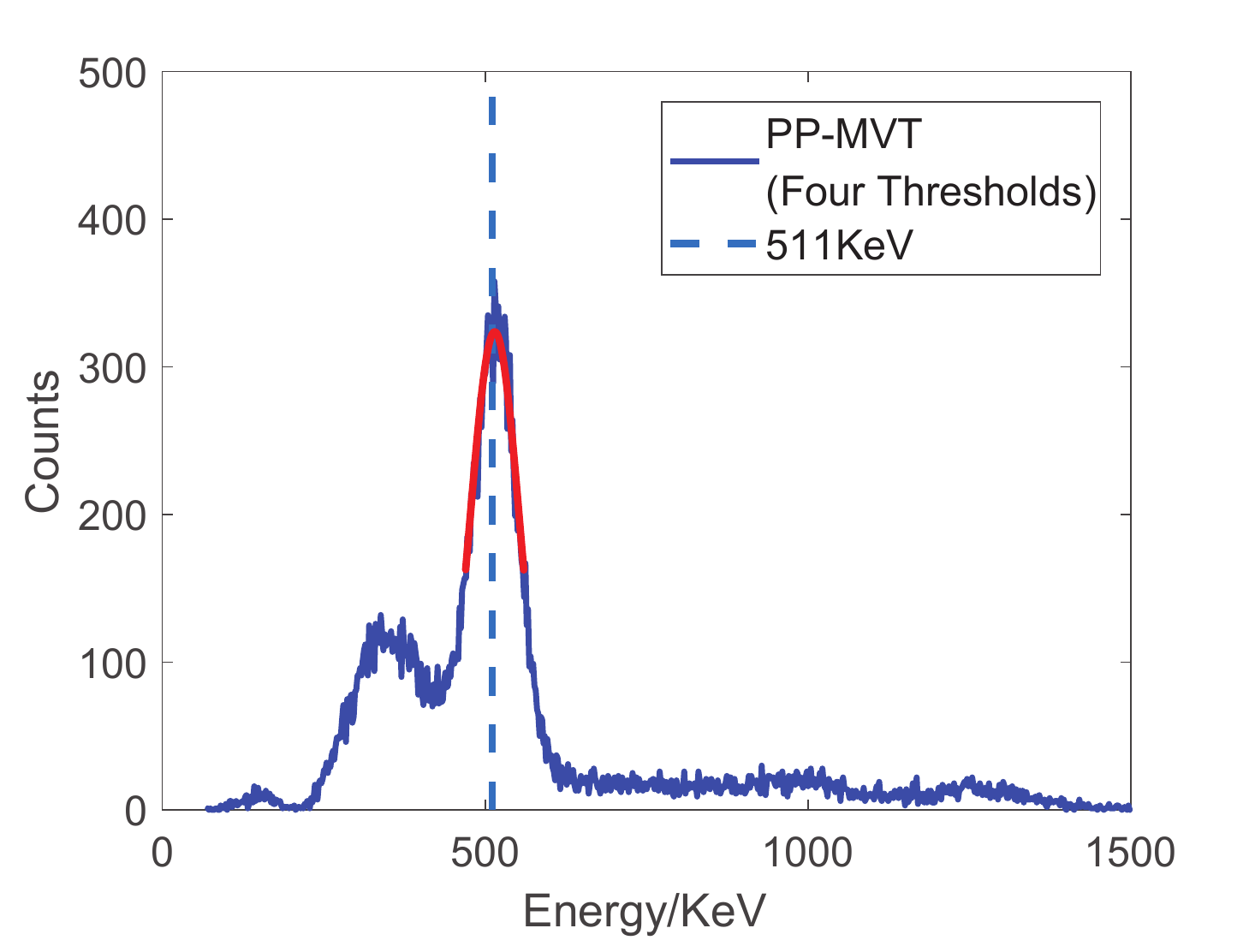}
\end{minipage}}}
\subfigure[]{
{\begin{minipage}[t]{0.45\textwidth}
\centering
\includegraphics[width=7cm]{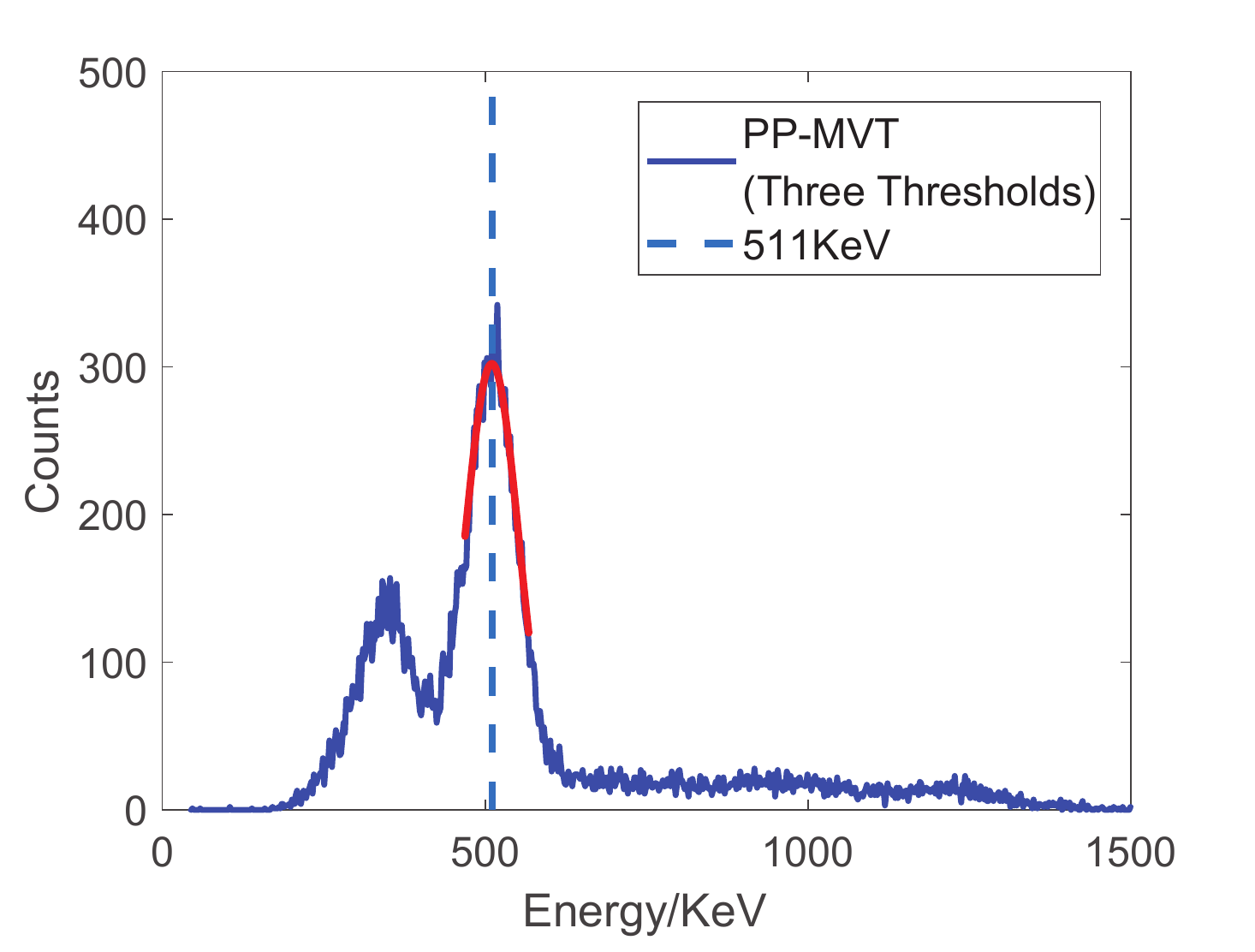}
\end{minipage}}}
\\
\subfigure[]{
{\begin{minipage}[t]{0.45\textwidth}
\centering
\includegraphics[width=7cm]{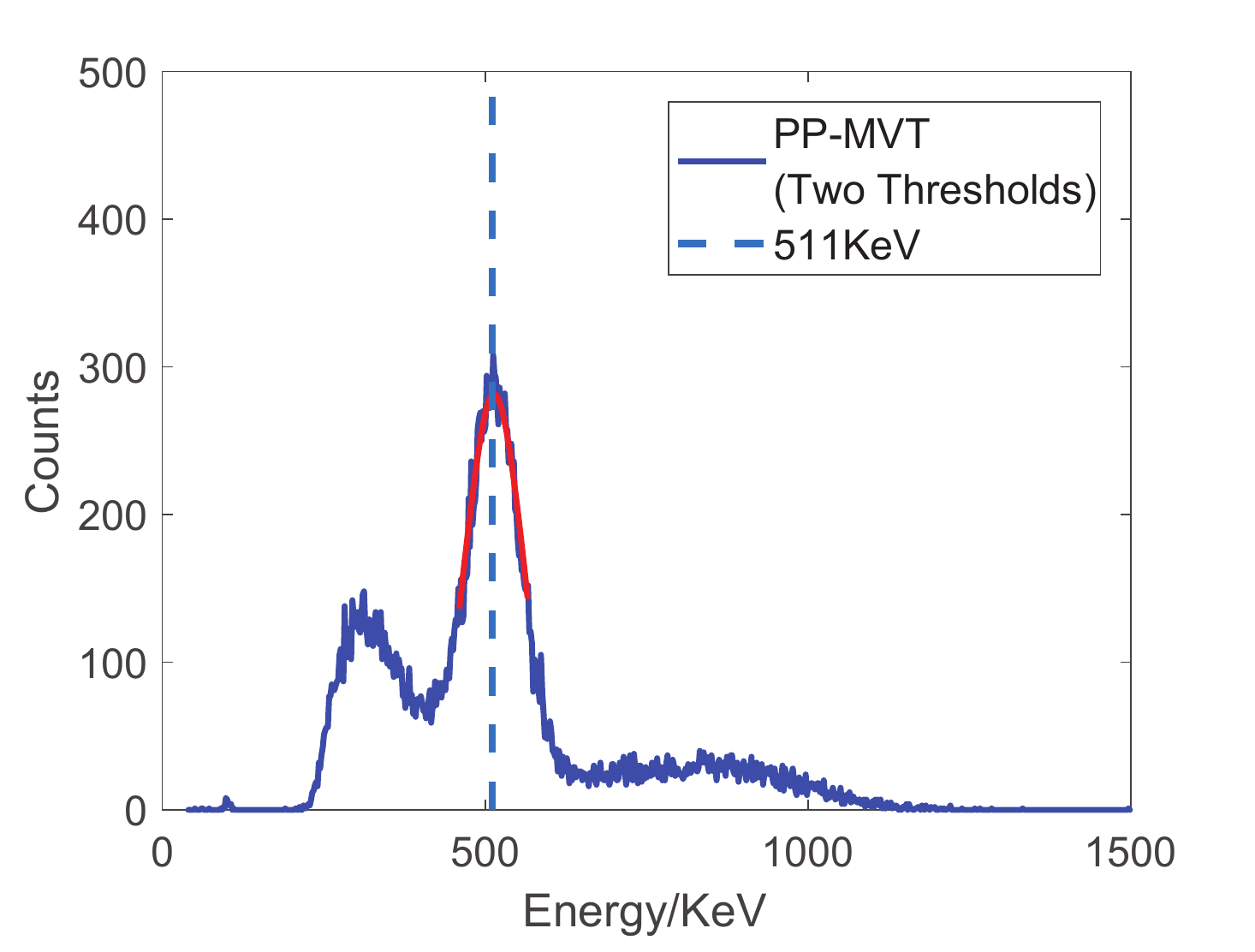}
\end{minipage}}}
\subfigure[]{
{\begin{minipage}[t]{0.45\textwidth}
\centering
\includegraphics[width=7cm]{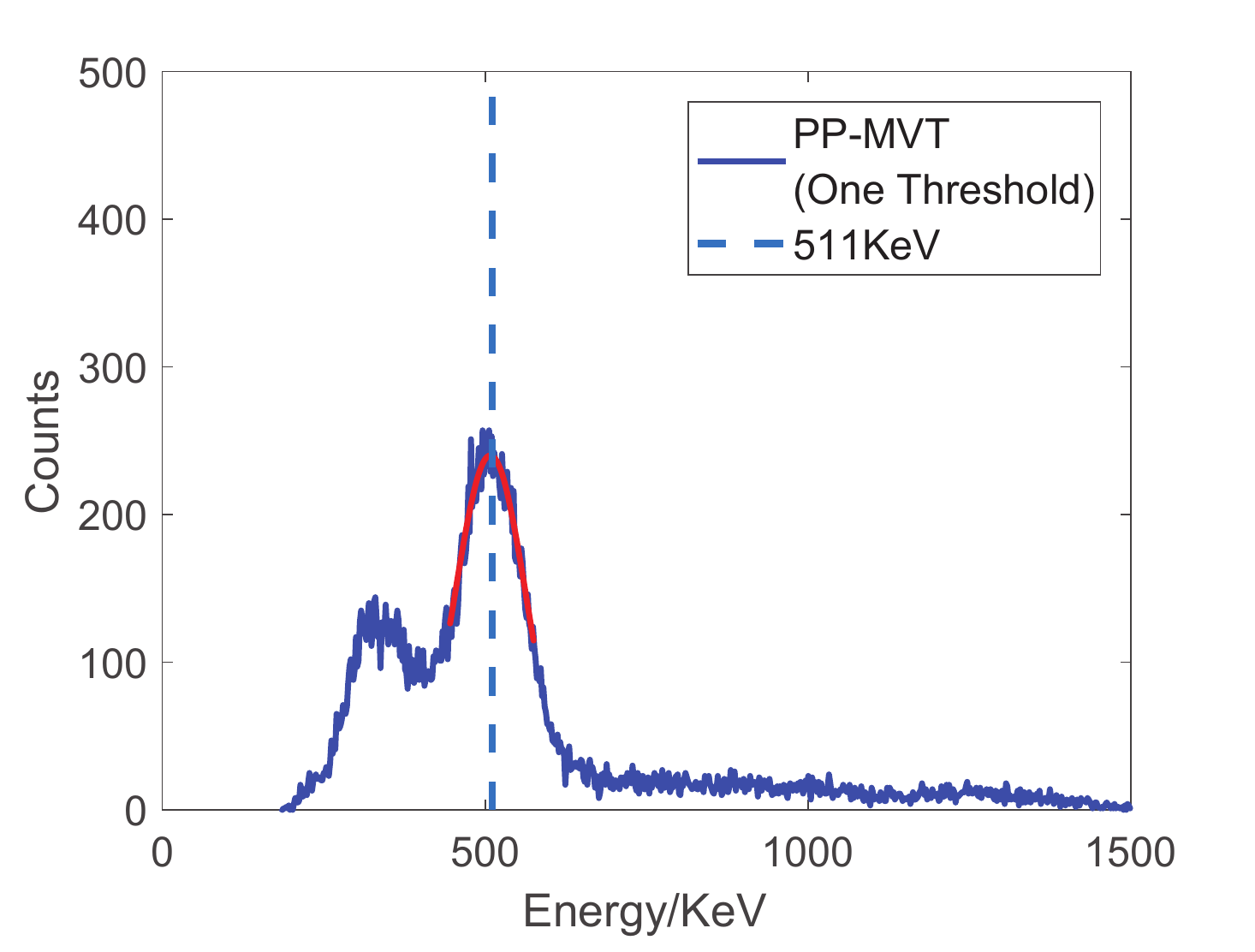}
\end{minipage}}}
\caption{Energy spectrum result for $^{22}Na$ single source detection. (a) The OSC result. (b) The traditional MVT result. (c) The PP-MVT result (four thresholds).  (d) The PP-MVT result (three thresholds). (e) The PP-MVT result (two thresholds). (f) The PP-MVT result (one threshold).\\
~\textbf{*}~The thick red curve indicates a Gaussian fit to the photopeak. }
\label{fig8}
\end{figure*}

Applying the traditional MVT and PP-MVTs of threshold number from four to one to process the 30,000 $^{22}Na$ pulses acquired by the OSC, the comparison result is shown in Fig.\ref{fig8} and Table\ref{tabI}. The energy resolution is calculated as the FWHM of the Gaussian fitting curve within an energy window of 450-650KeV width around the 511KeV photoelectric peak in the energy spectrum and the count number is the pulses number within this energy window. And when ignoring the nonlinearity correction of the SiPM, the energy resolution of OSC, MVT and PP-MVT with four thresholds could reach 10.27\%, 14.48\% and 13.11\% individually under the circumstance of a 450-650KeV energy window for this set of pulse data. The result of MVT is close to the one reported at Hua's work~\cite{33}.

\begin{table}[htbp]
\centering
\caption{results at 511 KeV for $^{22}Na$ single source detection, using one set of 30,000 pulses (energy window: 450-650 KeV).}
\small
\begin{flushleft}
\textbf{Energy resolution: $\%$ FWHM}\\
\textbf{Count: pulses number within the energy window}\\[1.2pt]
\end{flushleft}
\begin{tabular}{p{120pt}p{80pt}p{25pt}}
\toprule
Data Group    & Energy resolution & Count\\
\midrule
OSC           & 13.46$\% $ & 15259  \\
MVT (Four Thresholds)  & 19.94$\% $ & 14678  \\
PP-MVT (Four Thresholds)  & 17.50$\% $ & 15225  \\
PP-MVT (Three Thresholds)       & 19.35$\% $ & 15056  \\
PP-MVT (Two Thresholds)       & 20.48$\% $ & 14853  \\
PP-MVT (One Threshold)       & 25.46$\% $ & 14218  \\
\bottomrule
\end{tabular}
\label{tabI}
\end{table}

From the results, the PP-MVT surpasses the traditionaln MVT on energy resolution performance by 2.44\% and a 3.73\% ((15225-14678)/14678× 100\% ) increase on count number is gotten, when it involves the thresholds of the same amount as MVT; even PP-MVT removes one threshold, it still achieves a better energy resolution and a larger count number than MVT. And the trends of the changes on energy resolution and count number are generally synchronous: lower energy resolution is usually with a larger count number. For the PP-MVTs using multiple reconstruction models (with four, three and two thresholds), their count numbers are always larger than traditional MVT's. This indicates the importance of adaptive reconstruction models.

\section{Conclusions \& Discussion}

In this paper, we have proposed a Peak Picking Multivoltage Threshold digitizer and explored its potential advantages in energy determination in radiation measurement for PET. After stating the thresholds selection according to time or energy resolution, peak points acquirement by a designed circuit and the selection of adaptive reconstructing models through amplitude-deviation statistical analysis, the simulation experiment of PP-MVT shows a better energy resolution and a larger count number within the energy window of interest than traditional MVT when they adopt the same thresholds. Even being deprived of one threshold, the three-threshold PP-MVT(one peak point and six sampling points acquired by three thresholds) still achieves a better energy resolution and a bigger count number than the traditional MVT(eight sampling points acquired by four thresholds).

The improved PP-MVT has a better energy determination performance while maintaining the high count rate and low-cost advantages as avoiding fast ADCs. In the digital PET system, the use of PP-MVT can increase the accuracy and sensitivity in locating tinier lesions with limited cost demand. It can also reduce the dose amount of the radioactive tracers used, which helps prevent redundant radiation. Besides the molecular imaging technology, the proposed method could also be used in industrial field where emphasizes the ability to detect minor radiation or requires limited cost of a mess of signal conversion between electrical pulses and digital signal.

The limitation of PP-MVT also needs to be discussed: what further improvement the peak point could make on enhancing the energy resolution, and what influences the factors such as the number of the pulses could have on the performance of PP-MVT. Furthermore, more potential advantages of feature picked peak in PP-MVT should also be explored: how to use the amplitude feature voltage and amplitude arriving time to evaluate the performance of photoelectric converter or scintillation crystal, for example. It is possible that the research on mathematical theory and the application of computer technology (neural network/machine learning) could help solve these problems.

It should also be mentioned that the system tests using more sources in different environment need to be further done to strengthen the reliability and usability of PP-MVT. To actualize these tests, the sampling of peak point
should be presented in the engineering level, with problems like synchronizing the peak point sampling circuit and the original MVT circuits or finding duration counters of high precision to be solved. And to explain in theory, the mathematical explanation for verifying the method should also be provided in order to account for the deeper reasons for the introduction of the peak points. Therefore, we are currently implementing the proposed method in the engineering layer and considering extensive demonstration of the Peak Point Multi-voltage Threshold digitizer.

\acknowledgments

The authors want to present their sincere appreciation to Min Gao for her helpful discussions, Shuo Li and Chaoyong Tian for their selflessness on assisting experiment and sharing their valuable experience. The authors also want
to thank College Students' "InternetPlus" Innovation and Entrepreneurship Competition of China for initiating some parts of this work.

This work was supported in part by the Natural Science Foundation of China (NSFC) (Grants 61927801, 61425001, 61604059, 61601190 and 61671215);in part by National Training Program of Innovation and Entrepreneurship for Undergraduates (201810487061).


\section*{References}
\def\refname{}

\end{document}